\RequirePackage{ifpdf}
\RequirePackage{color}
\documentclass{JHEP3}

\usepackage{graphicx}
\usepackage{amssymb}
\usepackage[mathscr]{eucal}
\usepackage{amsmath}
\usepackage{cite}
\usepackage{afterpage}
\usepackage{slashed}
\usepackage{feynmp}

\newcommand{\gsimm}{\raise.3ex\hbox{$>$\kern-.75em\lower1ex\hbox{$\sim$}}}
\newcommand{\lsimm}{\raise.3ex\hbox{$<$\kern-.75em\lower1ex\hbox{$\sim$}}}

\newcommand{\be}{\begin{equation}}
\newcommand{\ee}{\end{equation}}
\newcommand{\ba}{\begin{eqnarray}}
\newcommand{\ea}{\end{eqnarray}}

\newcommand{\bea}{\begin{eqnarray*}}
\newcommand{\eea}{\end{eqnarray*}}

\title{The Speed of Galileon Gravity}
\author{Philippe Brax \\
Institut de Physique Th\'eorique, Universit\'e Paris-Saclay, CEA,CNRS,\\
F-91191Gif sur Yvette, France  \\ E-mail:
  \email{philippe.brax@cea.fr}}
 \author{Clare Burrage\\
 School of Physics and Astronomy, University of Nottingham, Nottingham, NG7 2RD, United Kingdom
  \\ E-mail:
  \email{Clare.Burrage@nottingham.ac.uk} }
\author{Anne-Christine Davis\\
  DAMTP, Centre for Mathematical Sciences, University of Cambridge,
  CB3 0WA, UK\\E-mail:
  \email{A.C.Davis@damtp.cam.ac.uk}}

\date{today}
\abstract{We analyse  the speed of gravitational waves in coupled Galileon models with an equation of state $\omega_\phi=-1$ now and a ghost-free  Minkowski limit. We find that the gravitational waves propagate much faster than the speed of light unless these models are small perturbations of cubic Galileons and the Galileon energy density is sub-dominant to  a dominant cosmological constant. In this case, the binary pulsar bounds on the speed of gravitational waves can be satisfied and the equation of state can be close to -1 when the coupling to matter and the coefficient of the cubic term of the Galileon Lagrangian are related. This severely restricts the allowed cosmological behaviour of Galileon models and we are forced to conclude that Galileons with a stable Minkowski limit  cannot account for the observed acceleration of the expansion of the universe on their own. Moreover any sub-dominant Galileon component of our universe must be dominated by the cubic term. For such models with gravitons propagating faster than the speed of light, the gravitons become potentially unstable and could decay into  photon pairs. They could also emit photons by Cerenkov radiation. We  show that the decay rate of such speedy gravitons into photons and the Cerenkov radiation are in fact negligible.
Moreover the time delay between the gravitational signal and light emitted by explosive astrophysical events could serve as a confirmation that a modification of gravity acts on the largest scales of the Universe. }

\begin{document}
\section{Introduction}
Gravitational waves have now been predicted for nearly a century and despite decades of experimental efforts, their existence is only confirmed by indirect evidence coming from the time drift of the period of binary pulsars. New experiments such as the advanced Laser Interferometry Gravitational-Wave Observatory (a-LIGO) \cite{Harry:2010zz}, the advanced VIRGO interferometer \cite{Accadia:2011zzc}, the Kamioka Wave Detector (KAGRA) \cite{Somiya:2011np}, the space based mission DECIGO \cite{Kawamura:2011zz} or eLISA \cite{AmaroSeoane:2012km}  will be able to test directly the existence of gravitational waves to improved levels. Gravity waves are also  important probes for theories going beyond Einstein's General Relativity (GR) \cite{Koyama:2015vza}. These theories are motivated by the discovery of the recent acceleration of the expansion of the Universe \cite{Copeland:2006wr} whose origin is still unknown. Models such as the quartic Galileons \cite{Nicolis:2008in} where a coupling between a scalar field and gravity is present predict a background dependent speed of gravitational waves.

In this work we focus on Galileon models \cite{Nicolis:2008in}.  These are a subset of the Horndeski action \cite{Horndeski:1974wa,Deffayet:2009mn} describing the most general scalar tensor model with second order equations of motion.  The Galileon terms on flat space are protected by a symmetry, the so called Galileon symmetry, which is softly broken on a curved spacetime background \cite{Deffayet:2009wt}.  In these models the cosmic acceleration is due to the presence of higher order terms in the derivatives compared to quintessence models where a non-linear potential, typically containing a term equivalent to a cosmological constant,  provides the required amount of vacuum energy. In vacuum the scalar mediates a fifth force of at least gravitational strength. Locally close to massive sources the scalar field is strongly influenced by matter and within the Vainshtein radius GR is restored. On cosmological time scales, the scalar field evolves. This cosmological time drift is  screened from matter fields whilst the average density of the universe is sufficiently high but  has consequences for the dynamics of gravity locally \cite{Chow:2009fm}. In particular the speed of gravitational waves in a massive environment is not protected from the evolution of the background cosmology by the Vainshtein mechanism \cite{Nicolis:2008in}, meaning that it can differ from the speed of light in a significant manner \cite{Jimenez:2015bwa}. We will review this calculation in Section \ref{sec:speedscreen}.

If we impose that the equation of state of the scalar field should be close to -1 now and the existence of a stable Minkowski limit of the theory in the absence of matter, both necessary conditions for a viable cosmology dominated by Galileons at late times and a meaningful embedding of the model in higher dimensions\footnote{We require this embedding in higher dimensional brane models with positive tension branes as a prerequisite first step towards a possible extension to fundamental theories such as string theory.} \cite{Brax:2014vla}, we find that the speed of gravitational waves would be much greater than one. This would increase the rate of emission of gravitational waves from binary pulsars.  As a result,  the speed of gravity in such a Galileon model is not compatible with the  bound that positive deviations of the speed of gravity from the speed of light cannot be more than one percent \cite{Jimenez:2015bwa,Hulse:1974eb}. We then conclude that  these Galileon models cannot lead to the acceleration of the Universe on their own and a certain amount of dark energy must be coming from a pure cosmological constant. This forces  the quartic Galileon terms to be subdominant to the cubic terms in order that  the binary pulsar bound can be satisfied. When this is the case, the time delay between gravity and light or even neutrinos can be as large as a few thousand years for events like the SN1987A supernova explosion. This would essentially decouple any observation of supernovae gravitational waves from the corresponding photon or neutrino signal coming from such explosive astrophysical events. On the other hand, a time difference as low as the uncertainty on the difference in emission time signal between neutrinos and gravity, e.g. up to $10^{-3}$ s for supernovae \cite{Nishizawa:2014zna}, would allow one to bound deviations of the quartic Galileon model from its cubic counterpart at the $10^{-14}$ level.

One possible caveat to these results would be if the superluminal gravitational waves do not reach our detectors because they either decay into two photons or lose all their energy through  Cerenkov radiation \cite{Kimura:2011qn}. We will show that  superluminal gravitational waves with a speed as large as one percent higher than the speed of light are not excluded by particle physics processes. A related possibility is at the origin of the stringent bounds on subluminal gravitational waves which could be Cerenkov radiated by high energy cosmic rays. As these high energy rays are observed the speed of gravitons cannot be significantly smaller than that of the particle sourcing the cosmic ray \cite{Moore:2001bv,Elliott:2005va}. We analyse the decay and the Cerenkov effect for superluminal gravitational waves and we find that their effects are negligible.

 Galileons have been widely studied both on  purely theoretical grounds, with results showing that this kind of models arise also in the context of massive gravity \cite{deRham:2011by} and braneworld models \cite{deRham:2010eu}. Constraints on the allowed cosmology of Galileon theories can be obtained from a wide variety of observations, unveiling a very rich phenomenology \cite{Appleby:2012ba, Chow:2009fm, Babichev:2011kq, Deffayet:2010qz, Mizuno:2010ag, Charmousis:2011bf, Barreira:2012kk, Barreira:2014jha, Silva:2009km, Kobayashi:2010wa, DeFelice:2010nf, DeFelice:2010as, Nesseris:2010pc, DeFelice:2010pv,Barreira:2013eea, Neveu:2013mfa}.  Here we consider for the first time the constraints that current and near future observations of gravitational waves can place on these theories.

In section \ref{sec:gal}, we recall details about Galileon models and show that quartic models with an equation of state close to -1 lead to very fast gravitons. In section \ref{sec:speedscreen}, we consider the influence of the Vainshtein mechanism on the propagation of gravity and we check that the screening mechanism does not protect the speed of gravity from large deviations compared to the speed of light. We also introduce models of subdominant Galileons whose gravitational waves have a speed which satisfies the binary pulsar bounds. In section \ref{sec:decay} we consider the decay rate of gravitons into two photons, and the Cerenkov radiation. We show that these processes are negligible for allowed differences between the speed of gravitons and photons. Finally In Section \ref{sec:delay} we discuss the time delay in the arrival time of gravitons and photons from explosive astrophysical sources. We conclude in section \ref{sec:conc}.

\section{Galileons}
\label{sec:gal}

\subsection{The Models}

In this paper, we are interested in models of modified gravity with a Galilean symmetry. They are potential candidates to explain the late time acceleration of the expansion of the Universe. They also lead to a modification of gravity on large scales.
Such Galileons are scalar field theories which have equations of motion that are at most second order in the derivatives. Moreover they are interesting dark energy candidates where an explicit cosmological constant is not compulsory.  Their Lagrangian reads in the Jordan frame defined by the metric $g_{\mu\nu}$
\begin{equation}
{\cal{L}} = \left(1+2 \frac{c_0 \phi}{m_{\rm Pl}}\right)\frac{R}{16\pi G_N}-\frac{c_2}{2}(\partial \phi)^2 -\frac{c_3}{\Lambda^3}\Box\phi (\partial \phi)^2 -\frac{c_4}{\Lambda^6}{\cal L}_4 -\frac{c_5}{\Lambda^9}{\cal L}_5 \;.\label{lag}
\end{equation}
 The common scale
\be
\Lambda^3 =H_0^2 m_{\rm Pl}
\ee
is chosen to be of cosmological interest as we focus on cosmological Galileon models which can lead to dark energy in the late time Universe.
We also require that $c_2>0$ to avoid the presence of ghosts in a Minkowski background. This theory could be rewritten in the Einstein frame where the conformal coupling of the scalar field to matter would be given by
 \be
A(\phi)=1+\frac{c_0\phi}{m_{\rm Pl}}
\ee
where $c_0$ is a constant.
The complete Galileon Lagrangian depends on operators with
  higher order terms in the derivatives  which are given by
 \begin{align}
 {\cal L}_4=&(\partial \phi)^2\left[2(\Box \phi)^2 -2 D_\mu D_\nu \phi D^\nu D^\mu \phi -R\frac{(\partial\phi)^2}{2}\right]\nonumber \\
 {\cal L}_5=& (\partial\phi)^2\left[(\Box\phi)^3 -3(\Box\phi)D_\mu D_\nu \phi D^\nu D^\mu \phi + 2 D_\mu D^\nu \phi D_\nu D^\rho\phi D_\rho D^\mu\phi\right.\\
&\left. -6 D_\mu\phi D^\mu D^\nu \phi D^\rho \phi G_{\nu\rho}\right].\nonumber
 \end{align}
These terms play an important role cosmologically. In the following and in the study of the cosmological evolution, we focus on the coupling of the Galileon to Cold Dark Matter (CDM) as the coupling to baryons is more severely constrained by the time variation
of Newton's constant in the solar system, at the one percent level, and does not play a significant role  for the background cosmology\cite{Brax:2015cla}.

This model is a subset of terms in the Horndeski action describing the most general scalar tensor theory with second order equations of motion
\ba
&&\mathcal{L}=K(\phi,X)-G_3(X,\phi) \Box\phi +G_4(X,\phi)R+G_{4,X}\left [ (\Box\phi)^2 -(D_\mu D_\nu \phi)^2\right] +\nonumber \\
&& G_5(X,\phi) G_{\mu\nu} D^\mu D^\nu \phi -\frac{1}{6} G_{5,X}\left [ (\Box \phi)^3 -3 \Box \phi (D_\mu D_\nu \phi)^2 +2(D^\mu D_\alpha \phi) (D^\alpha D_\beta \phi)(D^\beta D_\mu\phi)\right ]\nonumber
\ea
with the particular functions
\be
K= c_2 X, \ G_3(X)= -2\frac{c_3}{\Lambda^3} X,\ G_4(X,\phi)=\frac{A^2(\phi)}{16\pi G_N} + 2\frac{c_4}{\Lambda^6} X^2,\ G_5(X)= -6 \frac{c_5}{\Lambda^9} X^2
\ee
where $X= -\frac{(\partial \phi)^2}{2}$ is the kinetic energy of the field. In the following we shall focus on quartic Galileons with $c_5=0$ as this leads to both interesting cosmology and a non-trivial speed for gravitational waves.

\subsection{Cosmological Galileons}

We focus on the behaviour of Galileon models on cosmological scales in a Friedmann-Robertson-Walker background
\be
ds^2= a^2(-d\eta^2 +dx^2)
\ee
where $\eta$ is conformal time and we have set the speed of light $c=1$. The equations of motion of the Galileons can be simplified using the variable $x= \phi'/m_{\rm Pl}$
 where a prime denotes $'=d/d\ln a=- d/d\ln (1+z)$,  $a$ is the scale factor and $z$ the redshift. We define the scaled field $\bar y=\frac{\phi}{m_{\rm Pl} x_0}$,  the rescaled variables   $\bar x= x/x_0$ and $\bar H= H/H_0$ where $H$ is the Hubble rate, and the rescaled couplings \cite{Neveu:2013mfa}
$
\bar c_i= c_i x_0^i, \ i=2\dots 5, \ \ \bar c_0= c_0 x_0,\ \bar c_G= c_G x_0^2
$
where $x_0$ is the value of $x$ now. Notice that $x_0$ is not determined by the dynamics and is a free parameter of the model.  The cosmological evolution of the Galileon satisfies \cite{Appleby:2011aa}
\bea
\bar x'&=&-\bar x + \frac{\alpha\lambda -\sigma\gamma}{\sigma\beta-\alpha \omega}\label{eq:eom1} \\
\bar y'&=& \bar x \label{eq:eom2}\\
\bar H'&=& -\frac{\lambda}{\sigma} + \frac{\omega}{\sigma}\left(\frac{\sigma\gamma-\alpha\lambda}{\sigma\beta-\alpha\omega}\right)\label{eq:eom3}\\
\eea
where we have introduced the functions
\begin{align}
\alpha=& -3 \bar c_3 \bar H^3 \bar x^2 +15 \bar c_4 \bar H^5 \bar x^3+\bar c_0 \bar H +\frac{\bar c_2 \bar H \bar x}{6}  \label{eq:alpha} \\
\beta=& -2 \bar c_3 \bar H^4 \bar x +\frac{\bar c_2\bar H^2}{6} +9 \bar c_4\bar H^6 \bar x^2 \label{eq:beta} \\
\gamma=& 2\bar c_0 \bar H^2 -\bar c_3 \bar H^4 \bar x^2 +\frac{\bar c_2 \bar H^2 \bar x}{3}  \label{eq:gamma} \\
\sigma=& 2(1-2 \bar c_0 \bar y) \bar H -2 \bar c_0 \bar H \bar x +2 \bar c_3 \bar H^3 \bar x^3 -15 \bar c_4 \bar H^5 \bar x^4 \label{eq:sigma}\\
\lambda=&  3(1-2 \bar c_0 \bar y) \bar H^2 -2\bar c_0\bar H\bar x -2 \bar c_3 \bar H^4 \bar x^3+\frac{\bar c_2 \bar H^2 \bar x^2}{2}+\frac{\Omega_{r0}}{a^4}+\frac{15}{2} \bar c_4\bar H^6 \bar x^4\nonumber \\
\omega=&-2\bar c_0 \bar H^2 +2 \bar c_3\bar H^4 \bar x^2-12\bar c_4\bar H^6 \bar x^3.\label{eq:omega}
\end{align}
The Friedmann equation which governs the evolution of the Hubble rate can be written in a similar way
\be
(1-2 \bar c_0 \bar y)\bar H^2=\frac{\Omega_{m0}}{a^3}+\frac{\Omega_{r0}}{a^4} +2 \bar c_0 \bar H^2 \bar x+\frac{\bar c_2 \bar H^2 \bar x^2}{6}-2\bar c_3 \bar H^4 \bar x^3+\frac{15}{2} \bar c_4 \bar H^6 \bar x^4
\label{eq:friedman}
\ee
where the final four terms on the right hand side of Equation (\ref{eq:friedman})  correspond to the scalar energy density
\be
\frac{\rho_\phi}{H_0^2m_{\rm Pl}^2}=6 \bar c_0 \bar H^2 \bar x+\frac{\bar c_2 \bar H^2 \bar x^2}{2}-6\bar c_3 \bar H^4 \bar x^3+\frac{45}{2} \bar c_4 \bar H^6 \bar x^4
\ee
and the scalar pressure is
\begin{align}
\frac{p_\phi}{H_0^2m_{\rm Pl}^2}= &-\bar c_0[4\bar H^2 \bar x +2 \bar H (\bar H\bar x)']+\frac{\bar c_2}{2} \bar H^2 \bar x^2 +2 c_3 \bar H^3 \bar x^2 (\bar H \bar x)'  -\bar c_4[\frac{9}{2}\bar H^6 \bar x^4 +12 \bar H^6 \bar x^3\bar x'\nonumber \\
& +15\bar H^5 \bar x^4 \bar H']
\end{align}
from which we define the equation of state of dark energy
$
\omega_\phi=\frac{p_\phi}{\rho_\phi}
$
which must be close to -1 today if the Galileon is the dominant component of the universe at late times to comply with observational data. Normalising the field such $y_0=0$, which is a choice we can make without loss of generality,  the Friedmann equation gives one  constraint on the parameters of the model
\be
1= \Omega_{m0}+{\Omega_{r0}} +2 \bar c_0 +\frac{\bar c_2}{6}-2\bar c_3 +\frac{15}{2} \bar c_4
\label{con}
\ee
which is useful to reduce the dimension of the parameter space by one unit.

\subsection{The Speed of Gravitons}

\begin{figure}
\centering
\includegraphics[width=0.50\linewidth]{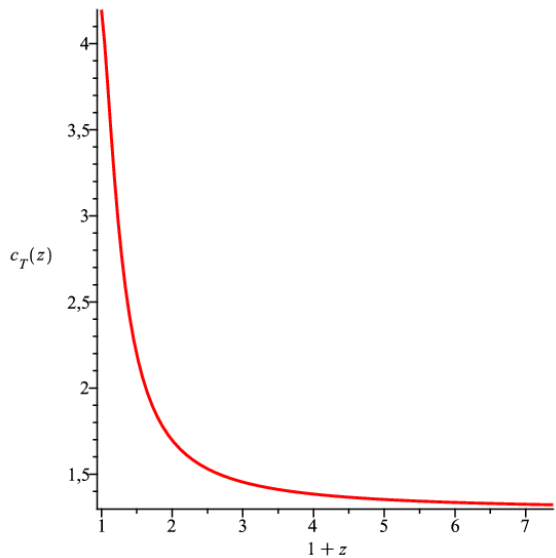}
\caption{The variation of the speed of gravitational waves as a function of redshift for a quartic Galileon model with an equation of state $\omega_\phi=-1$ now. The local constraints from binary pulsars rule out these types of models.}
\label{fig:speed}
\end{figure}

The speed of gravitational waves in a cosmological background is  given by \cite{Kimura:2011qn}
\be
c_T^2= \frac{\frac{A^2(\phi)}{16\pi G_N} + 2\frac{c_4}{\Lambda^6} X^2}{\frac{A^2(\phi)}{16\pi G_N} -6\frac{c_4}{\Lambda^6} X^2 }
\label{eq:speed}
\ee
where we restrict our analysis to the quartic Galileons  for simplicity. We will retrieve this result in the following sections where we study  the effect of the Vainshtein mechanism on the speed of gravitational waves.
In terms of cosmological quantities the speed of gravitons  is simply
\be
c_T^2= \frac{1 +2 \bar c_0 \bar y + \bar c_4 \bar H^4 \bar x^4}{{1 +2 \bar c_0 \bar y -3 \bar c_4 \bar H^4 \bar x^4}}
\ee
This sets the current speed to be
\be
c_{T0}^2= \frac{1+ \bar c_4}{1-3 \bar c_4}
\ee
When $c_4>0$, this is larger than one and no constraint from Cerenkov radiation of gravitons by cosmic rays applies. For the model with an equation of state $\omega_\phi=-1$ and $\bar c_2=1$ we have $\bar c_4\sim 0.3$ which implies that $c_T\sim 4$,  as shown in Figure \ref{fig:speed}.
Typically for the models with positive $c_2$ and an equation of state close to -1, the deviation of the speed of gravitational waves from one is far bigger than the percent level as allowed by the binary pulsar bound derived in \cite{Jimenez:2015bwa}.

\section{The Speed of Gravitons and Screening}
\label{sec:speedscreen}

\subsection{Screening Effects}

The speed of gravitons is tightly constrained by the drift of the period of binary pulsars. When the speed of gravitons exceeds the speed of light by more than one percent, the change in the period of binaries cannot accommodate observations \cite{Jimenez:2015bwa}. As we have seen, quartic Galileons with no ghosts in a Minkowski background, $\bar c_2=1$  and a cosmological equation of state now close to -1 have a cosmological speed which is much larger than the speed of light. One possible way out  which could reconcile both a large speed of gravitons on cosmological scales and a constrained one in the pulsar environment is the presence of screening in the form of the Vainshtein mechanism. We know that the Vainshtein mechanism suppresses the effects of the scalar on matter fields.  We now determine whether the same is true of gravitons. The speed of gravity depends on the Lagrangian
\be
{\cal L}_g= \sqrt{-g}[ G_4 R + G_{4X} ( (\Box \phi)^2 - D_\mu D_\nu \phi D^\nu D^\nu \phi)]
\ee
where we have restricted our model to quartic Galileons. We are interested in a near Minkowski geometry on the time  and spatial scales of the binary pulsars which are much smaller than the age and size of the Universe. First the Einstein Hilbert term is
\be
\sqrt{-g} G_4 R\supset -\frac{1}{4}G_4 (\partial_\mu h_{\nu\lambda}\partial^\mu h^{\nu\lambda})
\ee
where we have expanded
$g_{\mu\nu}= \eta_{\mu\nu}+ h_{\mu\nu}$ and used the transverse and traceless properties of gravitons $h^{\mu}_\mu=0, \ \partial_\mu h^\mu_\nu=0$ where indices are raised using the Minkowski metric. The term involving $G_{4X}$ can be evaluated using
\be
(\Box \phi)^2 - D_\mu D_\nu \phi D^\nu D^\nu \phi\supset \eta^{\mu\nu}\eta^{\alpha\beta}( D_\mu D_\nu \phi D_\alpha D_\beta \phi - D_\mu D_\alpha \phi D_\nu D_\beta \phi )
\ee
where $D_\mu D_\nu \phi = \partial_\mu\partial_\nu \phi - \Gamma^\lambda_{\mu\nu} \partial_\lambda \phi$. We are only interested in terms involving derivatives of $h_{\mu\nu}$ hence we have
\be
\Gamma^\lambda_{\mu\nu}=\frac{\eta^{\lambda\rho}}{2}( \partial_\nu h_{\mu\rho}+ \partial_\mu h_{\nu\rho}- \partial_\rho h_{\mu\nu})
\ee
and we notice that
\be
\eta^{\mu\nu} \Gamma^\lambda_{\mu\nu}=0
\ee
due to the traceless and transverse property. This implies that we are left with
\be
(\Box \phi)^2 - D_\mu D_\nu \phi D^\nu D^\nu \phi\supset -\eta^{\mu\nu}\eta^{\alpha\beta}\Gamma^\lambda_{\mu\alpha} \partial_\lambda \phi \Gamma^\rho _{\beta\nu} \partial_\rho \phi.
\ee
We are focusing on  waves $h_{ij}$ such that $k^i h_{ij}=0$ and $h^i_i=0$,
we can then use
\be
\Gamma^0_{ij}= \frac{1}{2} \partial_0 h_{ij},\ \Gamma^i_{jk}= \frac{1}{2} (\partial_j h_{ik}+ \partial_k h_{ij}-\partial_i h_{jk}), \ \ \Gamma^i_{0j}= \frac{1}{2} \partial_0 h_{ij}
\ee
and expand
\be
(\Box \phi)^2 - D_\mu D_\nu \phi D^\nu D^\nu \phi\supset -\eta^{\mu\nu}\eta^{\alpha\beta}( \Gamma^0_{\mu\alpha} \partial_0 \phi \Gamma^0 _{\beta\nu} \partial_0 \phi+
\Gamma^i_{\mu\alpha} \partial_i \phi \Gamma^j_{\beta\nu} \partial_j \phi+ 2\Gamma^0_{\mu\alpha} \partial_0 \phi \Gamma^i _{\beta\nu} \partial_i \phi
\ee
which becomes
\be
(\Box \phi)^2 - D_\mu D_\nu \phi D^\nu D^\nu \phi\supset - \Gamma^0_{i j} \partial_0 \phi \Gamma^0 _{ij} \partial_0 \phi
-\Gamma^i_{kl} \partial_i \phi \Gamma^j_{kl} \partial_j \phi+  2 \Gamma^i_{0k} \partial_i \phi \Gamma^j_{0k} \partial_j \phi - 2\Gamma^0_{jk} \partial_0 \phi \Gamma^i _{jk} \partial_i \phi.
\ee
In order to simplify the analysis, we consider the propagation of the gravitational waves along the spatial gradient of $\phi$, i.e. $ \partial_i \phi h^i_j=0$. This is in particular the case for spherical waves
with $k_0=\omega$ and $k_r=k\ne 0$ when $\phi$ depends only on $r$ and $t$. In this situation we assume that the time variation is coming from the background cosmological evolution and the radial dependence is sourced by an over-density of matter. We choose that $h_{ij}$ is  only  non-zero for $h_{\theta\theta}$ and $h_{\theta\phi}$. In this case we find that
\be
(\Box \phi)^2 - D_\mu D_\nu \phi D^\nu D^\nu \phi\supset - \frac{1}{4} (\partial_0 \phi)^2 (\partial_0 h_{ij})^2
-\frac{1}{4} (\partial^i \phi \partial _i h_{jk})^{{ 2}} +\frac{1}{2} \partial_0 \phi \partial_i \phi  \partial^i h_{jk} \partial_0 h^{jk}.
\ee
The kinetic terms resulting from ${\cal L}_g$ read
\be
{\cal L}_g \sim \frac{(\dot h_{ij})^2}{4}(G_4 - G_{4X}\dot \phi^2) - \frac{(\partial_i  h_{jk})^2}{4} G_4- \frac{(\partial^i \phi \partial_i  h_{jk})^2}{4}G_{4X}-\frac{G_{4X}}{2} \partial_0 \phi \partial_i \phi  \partial^i h_{jk} \partial_0 h^{jk}.
\ee
We now specialise to the case where the background scalar field is spherical and the gravitational wave is radial. We deduce the propagation equation, which we simplify by noting that  that the derivatives of $h_{ij}$ are much larger than the derivatives of the background field, (which also  implies that we can neglect mass terms for the graviton as they involve only derivatives of the background)
\be
\ddot h_{ij}( G_4 - G_{4X} \dot\phi^2) +2 G_{4X} \dot \phi \partial_r \phi \partial_r \dot h_{ij}  -\Delta h_{ij} (G_4+  (\partial_r \phi)^2 G_{4X})=0.
\ee
This can be easily analysed for spherical waves $h_{ij} \sim \frac{e^{i\omega t - k r}}{r}$ with variations on time and spatial scales much larger that the background ones,  leading to the dispersion relation
\be
\omega^2( G_4 - G_{4X} \dot\phi^2) -2 \omega k \dot \phi \partial_r \phi -k^2  (G_4+  (\partial_r \phi)^2 G_{4X})=0
\ee
In the absence of any spatial dependence $\partial_r \phi\equiv 0$, we retrieve that
\be
c_T^2=\frac{G_4}{G_4 - 2X G_{4X}}=\frac{\frac{m_{\rm Pl}^2}{2} +\frac{2 c_4 X^2}{\Lambda^6}}{\frac{m_{\rm Pl}^2}{2}-\frac{6 c_4 X^2}{\Lambda^6}}>1
\ee
if $c_4>0$, which is the result quoted in Equation (\ref{eq:speed}).
On the other hand when the spatial gradient dominates $\dot \phi^2 \ll (\partial_r \phi)^2$, we find that
\be
c_T^2 =1-\frac{ 2 X G_{4X}}{G_4 }=\frac{\frac{m_{\rm Pl}^2}{2} -\frac{2 c_4 X^2}{\Lambda^6}}{\frac{m_{\rm Pl}^2}{2}+\frac{2 c_4 X^2}{\Lambda^6}}<1
\ee
when $c_4>0$.
If the terms in $G_{4X}$ are negligible, the speed of gravitons is very close to one. In the quartic Galileon case, close to a spherical mass $M$, we have that $X$ is constant inside the Vainshtein radius and
\be
X=-\frac{1}{2} \Lambda^4 \left(\frac{c_{0b}M}{8\pi m_{\rm Pl} c_4}\right)^{2/3}
\ee
where $c_{0b}\lesssim 10^{-2}$ is the coupling of the Galileon to baryons \cite{Brax:2015cla}. For objects of masses of order of one solar mass with $c_4$ of order one, we find that $c_T$ is equal to one up to terms of order $X^2/m_{\rm Pl}^2 \Lambda^{6}$ which are very small of order $10^{-30}$. As a result the bounds on the speed of gravitons from cosmic rays are easily satisfied.
Unfortunately, the condition $\dot \phi^2 \ll (\partial_r \phi)^2$ is only valid when
\be
x_0^{{ 2}}\equiv \frac{\dot \phi_0^2}{m_{\rm Pl}^2 H_0^2}\ll (R_V H_0)^2.
\ee
As the Vainshtein radius $R_V$ must be less than the size of the horizon in order that there exist some effects of the Galileon on cosmological scales,
this requires a fine tuning of the initial conditions to have a slowly evolving Galileon now. When this is not satisfied, the speed of gravitons is not screened locally and it can deviate from one substantially, e.g. when $x_0\sim 1$. In fact the Vainshtein radius is given by
\be
R_V= \left(\frac{c_4^{{ 2 }} c_{0b}M}{(8\pi)^2 c_3^3 m_{\rm Pl} \Lambda^3}\right)^{1/3}
\ee
which is $R_V \sim 10^{-7} { H_0}$ for objects of one solar mass and $c_3\sim c_4\sim1$. This means that the effects of Vainshtein screening due to the presence of massive sources cannot be used to reduce the speed of gravitons to an acceptable level.  This was first realised in \cite{Jimenez:2015bwa}.  In the following we will avoid this fine tuning on the present time derivative of $\phi$, and make the speed of gravitational waves close to one by requiring  $c_4$ to be small, i.e. when the Galileon model is essentially cubic.

\subsection{Cubic Galileons}
If we assume that $x_0$ is not very small, screening does not modify the speed of gravitons and the speed of gravitational waves emitted by compact objects like binary pulsars can only be small when the influence of the quartic Galileon terms is negligible. For the purely cubic Galileons, the condition that the equation of state should be close to -1 implies that $c_2<0$  \cite{Barreira:2013eea}, a case that we discard as we require a well-defined Minkowski limit. We will see that one can preserve a positive $c_2$ and still impose that $c_4$ is small together with an equation of state close to -1 when the Galileon scalar field does not lead to all the dark energy of the Universe. Indeed if a dominant cosmological constant is added to the model,
 the dynamics can be integrated at late times and we have
\be
\dot\phi_0 \sim H_0 m_{\rm Pl} \sqrt{\frac{\Omega_m c_0}{c_3}}.
\ee
 This approximation is valid as long as the cubic term dominates over the quartic and quadratic ones, i.e. $ c_3 \frac{H\dot \phi}{\Lambda^3}\gg c_2$ and $c_4\frac{H\dot \phi}{\Lambda^3}\ll c_3$, this can be achieved when
\be
\bar c_2 \ll \sqrt{\Omega_m \bar c_0 \bar c_3},\ \ \bar c_4 \ll \sqrt{\frac{\bar c_3^3}{\Omega_m \bar c_0}}.
\ee
We then deduce that $X\sim \frac{H^2 m_{\rm Pl}^2}{2} \frac{\Omega_m c_0}{c_3}$ and finally we have that
\be
c^2_T \sim \frac{1+{c_4} \frac{X^2}{m_{\rm Pl}^2 \Lambda^6}}{1-{3c_4} \frac{X^2}{m_{\rm Pl}^2 \Lambda^6}}\sim 1+ { 4} \bar c_4 \Omega_m^2 \frac{\bar c_0^2}{\bar c_3^2}.
\ee
The deviation of the speed of gravity compared to one is small provided $\bar c_4$ is small enough.
As we have  generalised the Galileon models by requiring that only a fraction of the contents of the Universe is due to the Galileon, i.e. there is a cosmological constant on top
of the Galileon dark energy, the Friedmann equation is modified and the normalisation of the Hubble rate now  implies that
\be
\bar c_0 - \bar c_3= \frac{\Omega_g-\Omega_m-\Omega_r}{2} -\frac{\bar c_2}{6} -\frac{15\bar c_4}{4}
\ee
where
\be
\Omega_g= 1-\Omega_{\Lambda}
\ee
and $\Omega_\Lambda$ is the fraction of the contents of the Universe given by a pure cosmological constant.
When $\bar c_0 \sim \bar c_3 \gg \bar c_2$, we find that
\be
c_T^2 \sim 1 + {4} \bar c_4\Omega_m^2
\ee
and the binary pulsar constraint is satisfied provided
\be
\bar c_4 \lesssim  4\times 10^{-2}.
\label{eq:c4}
\ee
When $c_4$ is negative, the constraint from the propagation of cosmic rays is much stronger at the $10^{-17}$ level. As a result we will only focus  on the $c_4\ge 0$ case. Given the constraint of equation (\ref{eq:c4}) the cubic term is indeed dominant and we find that the dark energy equation of state now reads
\be
\omega_\phi= \frac{-3 (1-\Omega_g) + \bar p_\phi}{3 (1-\Omega_r-\Omega_m)}
\ee
where $\bar p_\phi = \frac{p_\phi}{H^2 m_{\rm Pl}^2}$ can be estimated to be $\bar p_\phi \sim -4 \bar c_0$. The equation of state is close to $-1$ provided we have
\be
\bar c_0 \sim \frac{3(\Omega_g-\Omega_r-\Omega_m)}{4}
\ee
meaning that $\bar{c}_0$ is a fixed function of the fraction of dark energy carried by the Galileon. This implies that $\bar c_2 \ll 1$ in order to guarantee that the cubic term dominates. These approximations are well verified numerically.

\section{Graviton Instability}
\label{sec:decay}

\subsection{Graviton Decay}

\begin{figure}
\centering
\begin{fmffile}{decay}
\begin{fmfgraph*}(150,105)
\fmfpen{thick}
\fmfleft{i1}
\fmfright{o1,o2}
\fmf{dbl_wiggly,label=$p$,label.dist=6thick}{i1,v1}
\fmf{photon,label=$p_1$}{v1,o1}
\fmf{photon,label=$p_2$}{v1,o2}
\fmfdotn{v}{1}
\end{fmfgraph*}
\end{fmffile}
\caption{A graviton decaying into two photons}
\label{fig:decay}
\end{figure}
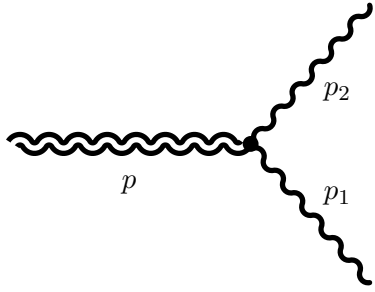

We have seen that the speed of gravitational waves emitted by binary pulsars can deviate from unity by a one percent for almost cubic Galileon models even if they are a subdominant component of the late universe.
In this case, the gravitons go faster than the speed of light and become unstable: they can decay into massless particles.  We focus here on the case in which the graviton decays into two photons as shown in Figure \ref{fig:decay}.
The interaction Lagrangian for this process reads
\be
{\cal L}_I= -\frac{\sqrt{-g}}{4} g^{\mu\nu}g^{\rho \sigma} F_{\mu\rho} F_{\nu\sigma}
\ee
where $F_{\mu\nu}=\partial_\mu A_\nu -\partial_{\nu} A_\mu$ is the field strength of photons. We consider the gravitons as transverse and traceless perturbations of the metric in a FRW background
\be
ds^2= a^2(\eta) ( \eta_{\mu\nu}+ h_{\mu\nu}).
\ee
The interaction Lagrangian has a part which leads to the decay of gravitons when $c_T>1$
\be
{\cal L}_I \supset \frac{1}{2} h^{\mu\nu} F_{\mu\lambda}F_\nu^\lambda
\ee
where indices have been raised using $\eta^{\mu\nu}$.
In the high energy regime where $k/a \gg H$, the graviton modes are given by
\be
h_k =\frac{1}{\sqrt{2\omega_k}}e^{-i\omega_k \eta}
\ee
where the dispersion relation satisfies
\be
\omega_k = c_s k
\ee
and the graviton field can be decomposed in terms of creation $a^{\lambda\dagger}_k$ and annihilation operators $a^{\lambda}_k $
\be
h_{\mu\nu}= \frac{1}{ a m_{\rm Pl}} \sum _\lambda \int \slashed{d}^3 k (\epsilon_{\mu\nu}^\lambda h_k e^{i\vec{k}.\vec{x}} a^{\lambda}_k + \bar \epsilon_{\mu\nu}^\lambda \bar h_k e^{-i\vec{k}.\vec{x}} a^{\lambda\dagger}_k)
\ee
where $\epsilon_{\mu\nu}^\lambda$ is the on-shell polarisation tensor of the graviton with $\lambda=\pm$ for its two polarisations. Similarly the photon field can be expanded as
\be
A_{\mu}=  \sum _\alpha \int \frac{\slashed{d}^3 k}{\sqrt{2k}} (\epsilon_{\mu}^\alpha  e^{-i\vec{k}.\vec{x}} b^{\alpha}_k + \bar \epsilon_{\mu}^\alpha  e^{i\vec{k}.\vec{x}} b^{\alpha\dagger}_k)
\ee
where $\epsilon_\mu^\alpha$ is the on-shell polarisation vector with $\alpha=\pm$ for the two photon polarisations. Here we have the 4d contraction $k.x= k_\mu x_\nu \eta^{\mu\nu}$.
The decay of the graviton of momentum $p$ into two photons is given by the integral
\be
\Gamma= \frac{1}{2c_T p} \int \frac{\slashed{d}^3 p_1}{2p_1} \frac{\slashed{d}^3 p_2}{2p_2 }\vert {\cal{M}}\vert^2 \slashed{\delta}^{(3)} (\vec{p}-\vec{p_1}-\vec{p_2})\slashed{\delta}(c_T p- p_1-p_2)
\ee
where the matrix element squared is simply
\begin{align}
\vert {\cal M}\vert^2= \frac{1}{4 a^2 m_{\rm Pl}^2}&\sum_{\alpha_1,\alpha_2,\lambda}
\vert (p_1.\bar \epsilon_2)(p_2.\epsilon_\lambda.\bar \epsilon_{\alpha_1}) +
(p_2.\bar \epsilon_1)(p_1.\epsilon_\lambda.\bar\epsilon_{\alpha_2})-(p_1.\epsilon_\lambda.p_2)(\epsilon_{\alpha_1}.\epsilon_{\alpha_2})\nonumber\\
&-(\epsilon_{\alpha_1}.\epsilon_\lambda.\epsilon_{\alpha_2})(p_1.p_2)\vert^2
\end{align}
where we have introduced the notation $a.\epsilon_\lambda.b= a_\mu \epsilon_\lambda^{\mu\nu} b_\nu$. We have also summed over the initial graviton polarisations.
Kinematically we find that
\be
(c_T^2 -1 ) p^2= 2p_1 p_2 (1-\cos\theta)
\ee
where $\theta$ is the angle between the two outgoing photons. We see that this process is only allowed when $c_T>1$. Moreover we find that the angle $\theta$ cannot be arbitrarily small but must satisfy
\be
\cos \theta < \frac{2}{c_T^2}-1
\ee
and the energy of the photons is such that
\be
\frac{p_1^2}{p^2} - c_T \frac{p_1}{p} + \frac{c^2_T-1}{2(1-\cos\theta)}=0.
\ee
 An order of magnitude for the decay rate can be obtained using ${\vert {\cal M}^2\vert \sim p_1^2 p_2^2/4 a^2 M_P^2}$. In this case we find that
\be
\Gamma\sim \frac{p^3 (c^2_T-1)}{16 \pi c_T m_{\rm Pl}^2{ a^2}} I
\ee
 where the phase space integral becomes
 \be
 I= \int dx_1 d\cos\theta \frac{x_1^3}{1-\cos\theta}\frac{1}{2x_1-c_T}={ \int_{c_T/2}^{ 1} dx_1\frac{x_1^2}{2c_T-x_1}}
 \ee
where $x_1=p_1/p$.
The integral is dominated by a collinear divergence $\cos\theta\to \frac{2}{c_T^2}-1$ when $c_T\to 1$ and $I\sim \ln (c^2_T-1)$. Finally we obtain that
 \be
{
\Gamma\sim \frac{p^3 (c^2_T-1) }{16 \pi c_T m_{\rm Pl}^2}\left(\frac{1}{8}(c_T-2)(5c_T+2) +c_T^2\ln\left(\frac{c_T}{2(c_T-1)}\right)\right)
}
\ee
which vanishes when $c_T\to 1$.
The number of gravitons $n$ satisfies the conservation equation
\be
\frac{d(a^3 n)}{d\eta}=-\Gamma a^3 n.
\ee
In cosmic time $dt= a d\eta$ and defining the physical momentum $p_{\rm phys}= \frac{p}{a}$, we find that
\be
{
\frac{dn}{dt}+3H n\sim - \frac{p_{\rm phys}^3 n (c^2_T-1) }{16 \pi c_T m_{\rm Pl}^2}\left(\frac{1}{8}(c_T-2)(5c_T+2) +c_T^2\ln\left(\frac{c_T}{2(c_T-1)}\right)\right).
}
\ee
For sources in our galactic environment and neglecting the dilution effect due to the expansion of the Universe, we find that the number of gravitons of momentum $p_{\rm phys} $ decays with a characteristic time
\be
{
\tau (p_{\rm phys})= \frac{16 \pi c_T m_{\rm Pl}^2}{p_{\rm phys}^3 (c^2_T-1)\left(\frac{1}{8}(c_T-2)(5c_T+2) +c_T^2\ln\left(\frac{c_T}{2(c_T-1)}\right)\right) }
}.
\ee
For astrophysical sources, this characteristic time far exceeds the age of the Universe and is not observable unless $c_T$ is fine tuned to be extremely  close to one.

\subsection{Cerenkov Radiation}

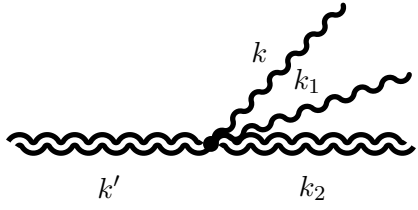
\begin{figure}
\centering
\begin{fmffile}{cerenkov}
\begin{fmfgraph*}(150,105)
\fmfpen{thick}
\fmfleft{i1}
\fmfright{o1}
\fmftop{t1,t2}
\fmfforce{(5w/6,h)}{t1}
\fmfforce{(w,3h/4)}{t2}
\fmf{dbl_wiggly,label=$k^{\prime}$,label.dist=6thick}{i1,v1}
\fmf{dbl_wiggly,label=$k_2$,label.dist=6thick}{v1,o1}
\fmf{photon,label=$k$,label.side=left,tension=0}{v1,t1}
\fmf{photon,label=$k_1$,label.side=left,tension=0}{v1,t2}
\fmfdotn{v}{1}
\end{fmfgraph*}
\end{fmffile}
\caption{A graviton Cerenkov producing two photons}
\label{fig:cerenkov}
\end{figure}

The gravitons can also emit two photons by the Cerenkov effect thereby losing energy and increasing the difficulty of detecting them. This process is shown in Figure \ref{fig:cerenkov}. In this case, the interaction Lagrangian contains
\be
{\cal L}_I \supset \frac{1}{16} h^2 F^2-\frac{1}{4} h^{\mu\nu} h^{\rho\sigma} F_{\mu\rho} F_{\nu\sigma} - \frac{1}{2} h^\mu_\sigma h^\sigma_\nu F^\mu_\delta F^{\nu \delta}
\ee
and the interaction Hamiltonian is
\be
H_I= \int d^3 x : {\cal L}_I :
\ee
where the operators in the Lagrangian are normal ordered. The emitted energy carried by the two photons is given by
\be
\langle E\rangle = 2 \sum_{\alpha} \int \slashed{d}^3 k k \langle b^\dagger_{k\alpha} b_{k\alpha}\rangle
\ee
where the averaged value is taken over the initial gravitons
\be
\langle  b^\dagger_{k\alpha} b_{k\alpha}\rangle = \langle \psi_\lambda \vert  b^\dagger_{k\alpha} b_{k\alpha}\vert \psi_\lambda\rangle
\ee
and the states are defined by
\be
\vert \psi_\lambda\rangle= \frac{1}{\sqrt 2} \int \slashed{d}^3 k \psi(k_1) a^\dagger_{k\lambda} \vert 0\rangle.
\ee
We normalise the states such that $\sum_\lambda \langle \psi_\lambda \vert \psi_\lambda\rangle =1$ implying that $\int \slashed{d}^3 k \vert \psi \vert^2=1$.
In perturbation theory and to second order the Cerenkov effect is obtained from
\be
\langle  b^\dagger_{k\alpha} b_{k\alpha}\rangle = 2 \Re \left ( \int_{-\infty}^\eta  d\eta_2 \int_{-\infty}^{\eta_2} d\eta_1 \langle  H_I(t_1)  b^\dagger_{k\alpha} b_{k\alpha}H_I(t_2))\rangle \right ).
\ee
Defining the polarisation tensors
\be
A_{\alpha\alpha'\lambda \lambda'}(p,p')= (\epsilon_\lambda . \bar \epsilon_{\lambda'})( (p.\epsilon_{\alpha'}) (p'.\bar \epsilon_{\alpha}) -(p.p')(\epsilon_\alpha.\bar \epsilon_{\alpha'}))
\ee
and
\begin{align}
B_{\alpha\alpha'\lambda \lambda'}(p,p')=&(p.\epsilon_\lambda.\bar \epsilon_{\lambda'}.\bar \epsilon_{\alpha'})(p'.\bar \epsilon_\alpha) + (p'.\epsilon_\lambda.\bar \epsilon_{\lambda'}.\bar \epsilon_{\alpha})(p'.\bar \epsilon_{\alpha'})- (p.\epsilon_\lambda.\bar\epsilon_{\lambda'}.p')(\bar \epsilon_\alpha.\bar \epsilon_{\alpha'})\nonumber\\
&- (p.p') (\bar\epsilon_\alpha .\epsilon_\lambda . \bar \epsilon_\lambda'.\bar\epsilon_\alpha')
\end{align}
where all the tensors are contracted according to their Lorentz indices and using the integral
\be
\left\vert \int_{-\infty }^\eta e^{i\omega \eta'} d\eta'\right\vert^2 = \eta \slashed{\delta}(\omega)
\ee
for large values of $\eta$, we find that the emitted energy per unit time is given by
\begin{align}
\frac{d\langle E\rangle }{d\eta}= \frac{1}{64 m_{\rm Pl}^4 a^4 c_T^4} \int & \slashed{d}^3 k \frac{\slashed{d}^3{k'}}{2k'}\frac{\slashed{d}^3{k_1}}{2k_1} \frac{\slashed{d}^3{k_2}}{2k_2}\\
&\times \sum_{\alpha_1,\alpha_2,\lambda'}
\vert \psi (k')\vert^2 \slashed{\delta}( c_T( k_2-k') +k_1 +k) \slashed{\delta}^3 ( \vec{k_2}-\vec{k} + \vec{k_1} + \vec{k} ) \vert {\cal M}\vert^2 \nonumber
\end{align}
where the matrix element squared is
\be
\vert {\cal M}\vert^2= (A_{\alpha_1\alpha_2\lambda'\lambda}(k,k_1)-4 B_{\alpha_1\alpha_2\lambda'\lambda}(k,k_1))(\bar A_{\alpha_1\alpha_2\lambda'\lambda}(k,k_1)-4 \bar B_{\alpha_1\alpha_2\lambda'\lambda}(k,k_1)).
\ee
The energy $k$ is the one of one tagged photon while the other one has an energy $k_1$. The initial graviton has momentum $k'$ and the outgoing one $k_2$. An estimate can be obtained using $\vert {\cal M}\vert^2\sim k^2 k_1^2$. We also take the initial graviton to have a peaked wave function at $k'=p$. The end result is that
\be
{
\frac{d\langle E\rangle }{d\eta}\sim \frac{(1-c_T^2)^2 }{ 2^{9}\cdot 3\cdot 5 \pi ^3c_T^4 a^4 }\frac{p^6}{m_{\rm Pl}^4}.
}
\ee
In cosmic time, the physical energy of the graviton decays according to
\be
{
\frac{dE_{\rm phys}}{dt} + H E_{\rm phys} \sim - \frac{(1-c_T^2)^2 }{2^{9}\cdot 3\cdot 5 \pi ^3 c_T^9 }\frac{E_{\rm phys}^6}{m_{\rm Pl}^4}
}
\ee
when $c_T\sim 1$.
The typical time of decay of the energy is
\be
{
\tau_{c}(E_{\rm phys}) =\frac{2^{9}\cdot 3\cdot 5 \pi ^3 c_T^9 }{(1-c_T^2)^2 }\frac{m_{\rm Pl}^4}{E_{\rm phys}^5}.
}
\ee
Again for astrophysical sources of gravitational waves, this time scale is longer than the age of the Universe.

\section{Time Delay}
\label{sec:delay}

The gravitons with a speed larger than the speed of light produced by astrophysical sources would arrive in our detector  well in advance of the light signal. Despite this, decays into two photons and Cerenkov radiation have a negligible effect on their propagation.
As a result, the difference with the speed of light or the speed of neutrinos  could affect the observations of both signals. An important possibility as we enter an era of multi-messenger astronomy\footnote{This was recently discussed for a difference choice of Horndeski scalar-tensor theory in \cite{Lombriser:2015sxa}.}. For instance, let us consider an explosive event such as the supernova SN1987A. In this case, the difference of emission times between neutrinos and gravitational waves is estimated to be around $10^{-3}$ s \cite{Nishizawa:2014zna}. For short gamma ray bursts, the emission times for photons and gravitational waves could differ by up to 500s \cite{Nishizawa:2014zna}. Typically we expect that gravitational waves would be reaching detectors earlier than neutrinos or photons by an amount
\be
\frac{\Delta t}{t} = \Delta c_T.
\ee
We have seen that current bounds from binary pulsars only constrain $\Delta c_T$ at the $10^{-2}$ level implying a  time delay, for sources one kpc away, of order 30 years. For the supernova SN19871A, gravitational waves could have reached the earth as early as 1700 years in advance. The tightest constraints on the difference of speed would come from supernovae around 1 kpc away with a time difference between neutrinos and gravitons greater than $10^{-3}$ s for $\Delta c_T \le 10^{-14}$. This is potentially twelve orders of magnitude lower than the binary pulsar bound. For Galileons, this would lead to an extraordinarily fine tuned model, which would behave like a cubic model, with the coefficient of the  quartic term suppressed by at least fourteen orders of magnitude.

\section{Conclusion}
\label{sec:conc}

We have analysed the behaviour of gravitational waves for  Galileon models that include quartic terms and have a stable Minkowski limit, and shown that only subdominant Galileon models where a significant part of the dark energy is due
to a cosmological constant can comply with the stringent binary pulsar bounds. When this is the case, the propagating gravitons do not suffer from particle physics instabilities such as  decay into two photons or Cerenkov radiation. As a result, the speed of gravitons remains superluminal but the difference between the speed of propagation of gravitons and photons cannot be more than one percent. In spite of this the time delay between the arrival of gravitational waves and light can be extremely large,  more than a thousand years for supernovae of the SN1987A type. More reasonable time delays can be expected for closer objects  when tighter bounds on the parameters of the models apply. The observation of such a time delay between the gravitational and light (or neutrino) signals coming from explosive astrophysical events would certainly be a hint that new physics requires a modification of GR on large scales.

\acknowledgments
P.B.
acknowledges partial support from the European Union FP7 ITN
INVISIBLES (Marie Curie Actions, PITN- GA-2011- 289442) and from the Agence Nationale de la Recherche under contract ANR 2010
BLANC 0413 01. CB is supported by a Royal Society University Research Fellowship.
 ACD acknowledges partial support from STFC under grants ST/L000385/1 and ST/L000636/1.

\end{document}